\documentclass[twocolumn]{IEEEtran}     

\usepackage{amsmath,epsfig}
\usepackage[T1]{fontenc}
\usepackage[utf8]{inputenc} 
\usepackage{psfrag}
\usepackage{array}
\usepackage{cite}
\usepackage{amsfonts}

\usepackage{amssymb}
\usepackage{color}
\usepackage{rotating}
\usepackage{subfigure}
\usepackage{graphicx}
\usepackage{float}
\usepackage{circuitikz}
\usepackage{tikz}
\usepackage{amsbsy} 
\usepackage{tcolorbox}
\usepackage{multirow}

\usepackage{acronym}
\usepackage{doi}

\def\blfootnote{\xdef\@thefnmark{}\@footnotetext}
\makeatother

\makeatletter
\def\blfootnote{\xdef\@thefnmark{}\@footnotetext}
\makeatother

\acrodef{ACC}[ACC]{average channel capacity}
\acrodef{AIC}[AIC]{Akaike information criterion}
\acrodef{CDF}[CDF]{cumulative distribution function}
\acrodef{EC}[EC]{ergodic capacity}
\acrodef{GoF}[GoF]{goodness-of-fit}
\acrodef{KS}[KS]{Kolmogorov-Smirnov}
\acrodef{KLD}[KLD]{Kullback-Leibler divergence}
\acrodef{ML}[ML]{maximum-likelihood}
\acrodef{MSE}[MSE]{mean-square-error}
\acrodef{MLE}[MLE]{maximum-likelihood estimation}
\acrodef{MTW}[MTW]{Multi-cluster Two-Wave}
\acrodef{OP}[OP]{outage probability}
\acrodef{PDF}[PDF]{probability density function}
\acrodef{RAD}[RAD]{resistor average distance}
\acrodef{SNR}[SNR]{signal-to-noise ratio}
\acrodef{UWB}[UWB]{ultra-wideband}

\makeatletter

\begin{document}

\title{\huge{How Should One Fit Channel Measurements to Fading Distributions for Performance Analysis?}}

\author{Santiago Fern\'andez, Jos\'e David Vega-S\'anchez, Juan E. Galeote-Cazorla and F. Javier L\'opez-Mart\'inez}

\maketitle
\begin{abstract}
Accurate channel modeling plays a pivotal role in optimizing communication systems, and 
fitting field measurements to stochastic models is crucial for capturing the key propagation features and to map these to achievable system performances. In this work, we shed light onto what's the most appropriate alternative for channel fitting, when the ultimate goal is performance analysis. Results show that \textcolor{black}{likelihood-based} and average-error metrics should be used with caution, since they can largely fail to predict outage probability measures. We show that supremum-error fitting metrics with tail awareness are more robust to estimate both ergodic and outage performance measures, even when they yield a larger average-error fitting.
\end{abstract}
%
\blfootnote{\noindent Manuscript received 12 December 2024. Revised XX XX. This work is supported by grant EMERGIA20-00297 funded by Consejería de Universidad, Investigación e Innovación (Junta de Andalucía), and by grant PID2023-149975OB-I00 (COSTUME) funded by MICIU/AEI/10.13039/501100011033 and FEDER/UE, and by predoctoral grant FPU22/03392.}
\blfootnote{\noindent S. Fern\'andez, J.E. Galeote-Cazorla and F.J. L{\'o}pez-Mart{\'i}nez are with Dept. Signal Theory, Networking and Communications, Research Centre for Information and Communication Technologies (CITIC-UGR), University of Granada, 18071, Granada, (Spain). Contact e-mail: $\rm santiago1988@ugr.es$.} 
\blfootnote{\noindent J. D.~Vega-S\'anchez is with	Colegio de Ciencias e Ingenier\'ias  ``El Polit\'ecnico", Universidad San Francisco de Quito (USFQ), Diego de Robles S/N, Quito (Ecuador) 170157.}
\blfootnote{\noindent This work has been submitted to the IEEE for publication. Copyright may be transferred without notice, after which this version may no longer be accessible}
\vspace{-4mm}
\section{Introduction}

\IEEEPARstart{P}{roperly} capturing the intrinsic nature of wireless propagation is vital in several aspects of practical wireless systems, such as performance optimization, efficient use of resources, or network planning and deployment, to name a few. This becomes particularly relevant in the context of ultra-reliable or low-latency communications, most notably in very low-outage operational regimes \cite{Eggers2019}. Specifically, in the context of channel modeling, a great body of generalized fading distributions have been proposed to characterize propagation mechanisms associated to new use cases in next-generation wireless networks \cite{RomeroJerez2017,Yacoub2016,Badarneh2020,Olyaee2023a}, and validated with field measurements in different scenarios \textcolor{black}{using distinct criteria}.

Taking a deeper look into the literature, there are different ways to carry out such empirical validation: one plausible alternative is through parameter estimation techniques, \textcolor{black}{with the simple} moment-based or the \textcolor{black}{nearly optimal} \ac{ML} estimators  \cite{Talukdar1991,Chen2005} implemented \textcolor{black}{from sample data}. Alternatively, channel fitting can be formulated as an optimization problem, i.e., finding the set of parameters that minimize a given error metric compared to the empirical sample distributions. Different \ac{GoF} criteria are used for this purpose: \ac{MSE} \cite{Abdi2002}, \ac{KLD} \cite{Kullback2012}, \ac{AIC} \cite{Yoo2017}, modified \ac{KS} \cite{RomeroJerez2017}, \ac{RAD} \cite{RAD}, and others \cite{Anjos2024,Gomes2023}; however, there is no clear consensus on what's the best universal criterion for fitting -- if any.

In communication theory, the distribution of fading ultimately determines the achievable performance of a communication system \cite{Simon1998}. Depending on the operational setting (ergodic or quasi-static) and other features (e.g., link adaptation, fading variation), the key metrics for system performance are the \ac{EC} and the \ac{OP} \cite{Lozano2012}. \textcolor{black}{Now, the key reason to map empirical channel measurements to a certain statistical distribution is to enable performance analysis of communication systems operating under such channels. However,} while the process of mapping sample data to a specific set of parameters and a given distribution is \textcolor{black}{quite common and well-investigated,} it \textcolor{black}{has been hitherto overlooked} whether the use of a certain \ac{GoF} criterion has an impact on the estimated performance. Hence, a key question emerges: \emph{--- what is the best channel fitting strategy when performance analysis comes into play?} {Through a number of experiments, we show that the \ac{GoF} criterion plays a relevant role when estimating the parameters of the distribution from the sample data, and that good \ac{GoF} values do not imply accuracy in performance evaluation. The use of tail-aware \ac{GoF} criteria provides improved robustness in the estimation of ergodic and outage performance measures, while conventional \ac{GoF} metrics that provide \textcolor{black}{likelihood-based or} average quality indicators often fail to predict \ac{OP} behaviors.
\vspace{-2mm}
\section{Channel fitting procedures}
\label{Sec:Fitting}
We consider a vector $\boldsymbol{\rm r}=\left \{ r_i \right \}_i^n$ with $i \in \left \{ 1,\dots,n \right \}$ consisting of $n$ samples representing the amplitude envelope. These can belong to empirical data retrieved from a measurement campaign, or be synthetically generated to represent a certain distribution.  
Following a model-based approach \cite{Molisch2005}, the fitting procedure defines some \ac{GoF} metric such that the optimal set of parameters of a target distribution that better approximates the sample in some sense is found. Mathematically, the underlying  distribution is fully characterized by an $M$-dimensional set of parameters collected in vector $\boldsymbol{\rm \lambda}=\left \{ \lambda_m \right \}_m^M$ with $m \in \left \{ 1,\dots,M \right \}$, where $\lambda_m$ denotes the $m$-th distribution's parameter. In the sequel, we denote $\hat{f_r} (r)$ and $\hat{F_r} (r)$ as the sample\footnote{They can be easily estimated from sample data, using for instance \texttt{ksdensity} and \texttt{ecdf} in \texttt{MATLAB}, respectively.} \ac{PDF} and \ac{CDF}, respectively. Similarly, $f_r (r;\boldsymbol{\rm \lambda})$ and $F_r (r;\boldsymbol{\rm \lambda})$ correspond to the target model \ac{PDF} and \ac{CDF}.

In the channel modeling literature, there are different alternatives to determine the \ac{GoF} when fitting statistical distributions to sample data. \textcolor{black}{For instance, one can use sample data to derive a log-likelihood metric (LLM) for a custom PDF, as}
\begin{equation}    \label{eqMLE}
    \mathcal{L}\left(\boldsymbol{r};{f}_r;\boldsymbol{\rm \lambda}\right) =  - \sum_{i = 1}^{n} \log\left({f}_r(r_i | \boldsymbol{\rm \lambda})\right),
\end{equation}
\textcolor{black}{with $\log$ being the natural logarithm. Then, the set of parameters that maximizes \eqref{eqMLE} provides the ML estimator. Alternative} approaches based on \ac{MSE} measures have the form \cite{Gomes2023}
\begin{equation}\label{eq1}
        \epsilon_{\mathrm{MSE}}\left(\hat{f}_r,{f}_r;\boldsymbol{\rm \lambda}\right) \triangleq \frac{1}{\mathcal{K}}\sum_{k=1}^{\mathcal{K}}(\hat{f}_{ r}(x_k)-f_{ r}(x_k;\boldsymbol{\rm \lambda}))^2,
\end{equation}
\noindent where $\mathcal{K}$ is the number of points on which the empirical \ac{PDF} is estimated, and $x_k$ correspond to the abscissa coordinates for such estimations.  
Alternative formulations of this metric can include normalization factors, root-based definitions, or be computed over the \ac{CDF} \cite{Nikookar1993} or log-\ac{PDF} \cite{Anjos2024}. In the latter case, \eqref{eq1} has the form of a sum of log-likelihood ratios (LLRs).

The \ac{KLD} is probably the most popular information-theoretic distance measure, defined as 
\begin{equation}
\label{eq2}
    \epsilon_{\rm{KLD}}\left(\hat{f}_r,f_r;\boldsymbol{\rm \lambda}\right) \triangleq \frac{1}{\mathcal{K}} \sum_{k=1}^\mathcal{K}  \hat{f}_r (x_k) \log \left( \tfrac{\hat{f}_r (x_k)}{{f_r} (x_k;\boldsymbol{\rm \lambda})}\right),
\end{equation}
\noindent where the metric also has the form of a sum of LLRs, now scaled by the sample probability density values at $x_k$. Since the \ac{KLD} is not symmetric by construction (i.e., the sample and target distributions are not interchangeable), a symmetrized definition is often used \cite{RAD} through the \ac{RAD}:
\begin{equation}\label{eq3}
    \epsilon_{\rm{RAD}}(\boldsymbol{\rm \lambda}) \triangleq \left( \tfrac{1}{\epsilon_{\rm{KLD}}\left ( f_r ,\hat{f_r} \right )} + \tfrac{1}{\epsilon_{\rm{KLD}}\left ( \hat{f_r} ,f_r  \right )}\right)^{-1}.
\end{equation}

Finally, a modified \ac{KS} statistic \cite{RomeroJerez2017} is also used in the literature when fitting the empirical \ac{CDF} to data in logarithmic scale. Mathematically, it is defined as the supremum of the distance between the sample and target \ac{CDF}, i.e.,
\begin{equation}\label{eq4}
    \epsilon_{\rm{KS}}(\boldsymbol{\rm \lambda}) \triangleq  \sup_{ x_k } |\log_{10} \hat{F_r} (x_k) - \log_{10} F_r (x_k;\boldsymbol{\rm \lambda})|.
\end{equation}

 In \eqref{eq4}, log-\ac{CDF} is used to outweigh the fit at the channel distribution's left tail (i.e., amplitude values close to zero). In this region, the fading is more severe and such events ultimately determine the error performance of the system. 

 \textcolor{black}{The \emph{likelihood-based} GoF criterion in \eqref{eqMLE} is computed by evaluating the target PDF in the sample data values. }By construction, \textcolor{black}{\emph{average-error}} \ac{GoF} criteria in \eqref{eq1}-\eqref{eq3} are calculated by averaging out over the empirical distribution. Conversely, \textcolor{black}{\textit{supremum-error} GoF in} \eqref{eq4} represents a worst-case metric (i.e., the maximum difference), with tail awareness as an additional feature. 
\vspace{-3mm}
\section{Methods}
\label{Sec:Methods}
The goal is to determine whether the \ac{GoF} criterion used for fitting has an impact on the system achievable performance. For this purpose, we conduct a number of experiments, which can be described as follows:
\begin{enumerate}
    \item Estimate the empirical distributions (i.e., either \ac{PDF} or \ac{CDF}) of the signal amplitude from sample data. 
    \item Find the sets of parameters $\boldsymbol{\rm \lambda}_j$ of the target distribution that optimizes the \ac{GoF} metric $j$, for $j\in\left\{\rm ML, MSE, RAD, KS\right\}$.
    \item For each $\boldsymbol{\rm \lambda}_j$, compute the \ac{GoF} metric obtained for the remaining criteria.
    \item For each $\boldsymbol{\rm \lambda}_j$, compute a set of benchmark performance metrics and compare to those obtained by using the sample distributions.
\end{enumerate}

To carry out these experiments, we define a set of performance metrics for benchmarking purposes, and a reference channel model for the target distribution used for fitting.
\vspace{-3mm}
\subsection{Benchmark Performance Metrics}
\label{Sec:PM}
Let us consider a canonical wireless communication system, where the \ac{SNR} at the receiver side is expressed as $\gamma=\overline\gamma\frac{r^2}{\Omega}$, with $r$ being the amplitude envelope, $\overline\gamma$ the average \ac{SNR}, and $\Omega=\mathbb{E}\left\{r^2\right\}$, and $\mathbb{E}\left\{\cdot\right\}$ being the expectation operator.

When channel state information is available at the receiver side only, the \ac{EC} (per bandwidth unit) is obtained by averaging the instantaneous capacity over all fading states, as
\begin{equation}\label{eq5}
    \overline{C}\left[\rm bps/Hz\right] = \int_0^\infty \log_2(1+x) f_{\gamma} (x) \mathrm{d} x,
\end{equation}
\noindent where $f_{\gamma} (\cdot)$ denotes the \ac{PDF} of $\gamma$, which can be derived from that of $r$ by a simple transformation.

The \ac{OP} is defined as the probability that the instantaneous channel capacity falls below a threshold rate $R_{\rm th}$, i.e., \cite{Simon2005}
\begin{align}\label{eq6}
    P_{\rm out}  & = \Pr \{ \log_2(1+\gamma) < R_{\rm th} \} = F_{\gamma} \left(\gamma_{\rm th} \right), 
\end{align}
\noindent where $F_{\gamma} (\cdot)$ is the \ac{CDF} of $\gamma$ and $\gamma_{\rm th}=2^{R_{\rm th}} - 1$. 
\begin{table*}[ht!]
\caption{\small \ac{GoF} results for Experiments 1 and 2. \textcolor{black}{LLM values $\mathcal{L}$ are normalized to sample size $n$ for range reduction.}}
\vspace{-2mm}
{\tiny
        \hspace{-7mm}
	\begin{tabular}
    {cc|c|c|c|c||c|c|c|c|l}
		\cline{3-10}
		& & \multicolumn{4}{ c|| }{\fcolorbox{black}{gray!65}{\textbf{\scriptsize{Experiment 1: Synthetic data}}}} & \multicolumn{4}{ c| }{\fcolorbox{black}{gray!65}{\textbf{\scriptsize{Experiment 2: Field Measurements}}}}\\ \cline{3-10}
		& & \multicolumn{4}{ c|| }{\textbf{\scriptsize{Optimization criteria}}} & \multicolumn{4}{ c| }{\textbf{\scriptsize{Optimization criteria}}}\\ \cline{3-10}
		& & \textbf{ML} & \textbf{MSE} & \textbf{RAD} & \textbf{KS} & \textbf{ML} & \textbf{MSE} & \textbf{RAD} & \textbf{KS}\\ \cline{1-10} 
		\multicolumn{1}{ |c}{\multirow{4}{*} {{\rotatebox[origin=c]{90}{\textbf{\ac{GoF}}}}}} &
		\multicolumn{1}{ |c| }
		{$\mathcal{L}$} & \fcolorbox{black}{gray!35}{$-2.789 \times 10^{-1}$} & $-2.790 \times 10^{-1}$ & $-2.790 \times 10^{-1}$ & $-2.929 \times 10^{-1}$ &   \fcolorbox{black}{gray!35}{$-3.529 \times 10^{-1}$} & $-3.542 \times 10^{-1}$ & $-3.541 \times 10^{-1}$ & $-3.628 \times 10^{-1}$ & \\ \cline{2-10}
		\multicolumn{1}{ |c  }{}                        &
		\multicolumn{1}{ |c| }{$\epsilon_{\rm MSE}$} & $1.518 \times 10^{-4}$ & \fcolorbox{black}{gray!35}{$9.472 \times 10^{-6}$} & $9.820 \times 10^{-6}$ & $6.663 \times 10^{-3}$ & $4.403 \times 10^{-4}$ & \fcolorbox{black}{gray!35}{$2.416 \times 10^{-4}$} & $2.439 \times 10^{-4}$ & $4.073 \times 10^{-3}$ &    \\ \cline{2-10}
		\multicolumn{1}{ |c  }{}                        &
		\multicolumn{1}{ |c| }{$\epsilon_{\rm RAD}$} & $3.914 \times 10^{-3}$ & $1.222 \times 10^{-3}$ & \fcolorbox{black}{gray!35}{$1.204 \times 10^{-3}$} & $3.150 \times 10^{-2}$& 
		$7.723 \times 10^{-3}$ & $5.759 \times 10^{-3}$ & \fcolorbox{black}{gray!35}{$5.695 \times 10^{-3}$} & $2.274 \times 10^{-2}$ &    \\ \cline{2-10}
		\multicolumn{1}{ |c  }{}                        &
		\multicolumn{1}{ |c| }{$\epsilon_{\rm KS}$} & $4.985 \times 10^{-1}$ & $1.865 $ & $1.887 $ & \fcolorbox{black}{gray!35}{$1.611 \times 10^{-1}$} & 
		$1.905 $ & $1.733 $ & $1.728 $ & \fcolorbox{black}{gray!35}{$8.616 \times 10^{-2}$} &    \\ \cline{1-10}
		\multicolumn{1}{ c  }{} & \multicolumn{1}{ |c|}{\multirow{3}{*} {$\boldsymbol{\lambda}$}}                       &
		$\Big[22.3297; \hspace{0.5cm}$ & $\Big[11.4531; \hspace{0.5cm}$ & $\Big[11.7369;\hspace{0.5cm}$ & $\Big[7.8257;\hspace{0.5cm}$ & 
		$\Big[0.8662; \hspace{0.5cm}$ & $\Big[0.3786; \hspace{0.5cm}$ & $\Big[0.3558;\hspace{0.5cm}$ & $\Big[9.5639;\hspace{0.5cm}$ &    \\ 
		\multicolumn{1}{ c  }{} & \multicolumn{1}{ |c|}{\multirow{3}{*} {}}                       & $0.8818;$ & $0.9157;$ & $0.91402; $ & $0.87751; $ &     
		$0.0005;$ & $0.0016; $ & $0.1332; $ & $0.6910;$ &    
		\\ 
		\multicolumn{1}{ c  }{} & \multicolumn{1}{ |c|}{\multirow{3}{*} {}}                       &
		$\hspace{0.5cm} 1.4190\Big]$ & $\hspace{0.5cm} 2.3482\Big]$ & $\hspace{0.5cm} 2.2934\Big]$ & $\hspace{0.5cm} 1.9973\Big]$ &  
		$\hspace{0.5cm} 1.5735\Big]$ & $\hspace{0.5cm} 1.8706\Big]$ & $\hspace{0.5cm} 1.8733\Big]$ & $\hspace{0.5cm} 0.7430\Big]$ &    \\ \cline{2-10}
	\end{tabular}
 }      \label{tab:GoF_Synthetic_Empirical}
\end{table*}
\vspace{-3mm}
\subsection{Baseline Channel Model}
\label{Subsec:RCM}
While any statistical distribution may be used to approximate the behavior of wireless channels, only a subset of distributions that comply with propagation laws can efficiently capture the real nature of these channels, desirably through a reduced set of physically meaningful parameters. Among the vast number of fading distributions in the literature, we decided to stick to the \ac{MTW} model recently proposed in \cite{Olyaee2023a}, since it captures relevant propagation phenomena (dominant and diffuse components, and clustering) through only three shape parameters. 

The \ac{MTW} model applies in a general scenario on which the resulting signal comes from the combination of a certain number of clusters, denoted by $\mu$ \cite{Yacoub2007}. Within any of these clusters, the signal is formed by $L_{i}$ dominant specular components plus a diffuse component \cite{Romero2022}. Then, the complex baseband signal amplitude of the \emph{i-th} cluster can be expressed as
\begin{equation} 	
Z_i  = \sum_{\ell=1}^{L_i} V_{i,\ell}  \exp \left( {j\phi _{i,\ell} } \right) + X_i + jY_i, 
\label{eq:001} 	
\end{equation} 
\noindent where $ V_{i,\ell} $ and $\phi _{i,\ell}$ are the amplitude and phase for the \emph{$\ell$}-th specular component, being the latter uniformly distributed in $[0, 2\pi)$, i.e., $\phi _{i,\ell}  \sim \mathcal{U}[0,2\pi)$. The per-cluster diffuse component is captured through $(X_i + jY_i)$, as a complex Gaussian random variable (RV) with $X_i,Y_i \sim \mathcal{N}(0,\sigma^2_i)$. 

Under the conventional assumption that the delay-time spreads of the different clusters are large enough so that they can be individually detected and combined at the receiver \cite{Yacoub2007}, we can express the squared amplitude envelope as $r^2=\sum_{i=1}^\mu { |Z_i|^2}$. The three-parameter \ac{MTW} model emerges when $L_1=2$, $L_{i>1}=0$, and $\sigma_i^2=\sigma^2$, so that the set of defining parameters is $\boldsymbol{\rm \lambda}=\left \{ K, \Delta,\mu \right \}$, with $\mu>0$, $K \triangleq \frac{V_1^2+V_2^2 }{2\mu\sigma^2}\geq 0,
$ and $\Delta=\frac{2V_1V_2}{V_1^2+V_2^2}\in[0,1]$. For completeness, the scale parameter is given by $\Omega=\mathbb{E}\left\{r^2\right\}=V_1^2+V_2^2+2\mu\sigma^2$.
\vspace{-3mm}
\section{Experiments}
Our goal is to determine the most suitable fitting strategy to ensure accuracy in performance analysis evaluation. 
So, we conduct two different experiments for this purpose: 
\begin{enumerate}
    \item {\bf Experiment 1 (E1)}: We generate a large sample of synthetic data from a known distribution, corresponding to the reference channel model in \ref{Subsec:RCM} with parameters $\boldsymbol{\rm \lambda}_{\rm E1}$. Then, we follow the procedure described in Section \ref{Sec:Methods} using the MTW distribution as the target model.
    \item {\bf Experiment 2 (E2)}: \textcolor{black}{
    We use empirical data from a measurement campaign in the W-band ($75$-$110$ GHz), carried out in one of CITIC-UGR's laboratories, and follow the subsequent steps in Section \ref{Sec:Methods}, again using the MTW distribution as the target model.}   
\end{enumerate}

Since E1 is conducted using a finite sample of synthetic data belonging to a known distribution, the estimated set of parameters $\boldsymbol{\rm \lambda}_j$ is expected to be close to the true set of parameters $\boldsymbol{\rm \lambda}_{\rm E1}$. However, the empirical \ac{CDF} may deviate from the theoretical \ac{CDF} evaluated from $\boldsymbol{\rm \lambda}_{\rm E1}$ in the vicinity of $0^{\rm +}$ because of the finite data set. In the case of E2, the true distribution is unknown. For each of the experiments, the GoF results are summarized in Table \ref{tab:GoF_Synthetic_Empirical}, together with the set of parameters $\boldsymbol{\rm \lambda}_j=\left[K_j;\Delta_j;\mu_j\right]$ that \textcolor{black}{maximizes (ML) or }minimizes (MSE, RAD, KS) the GoF metric $j$. The optimization mechanisms have been implemented with \texttt{MATLAB} in-built routine\textcolor{black}{s \texttt{mle} and} \texttt{fminsearch}. These experiments are discussed next. 
\begin{figure*}[ht!]
    \centering
    \subfigure[\ac{PDF} vs. $r$.] {\includegraphics[width=0.3\textwidth]{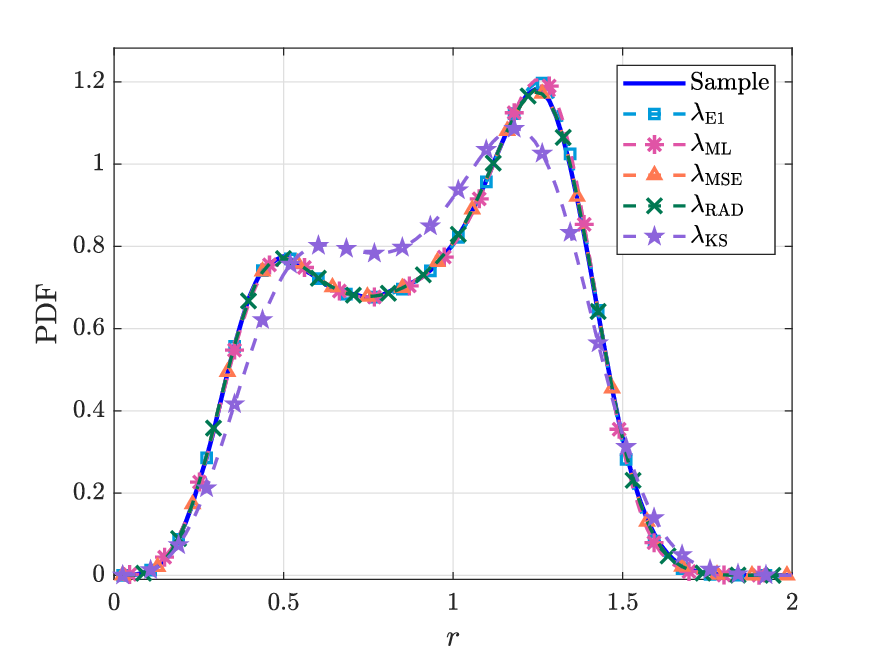} \label{fig:Synth_PDF}} 
    \subfigure[\ac{CDF} vs. $r$ (dB).]
    {\includegraphics[width=0.3\textwidth]{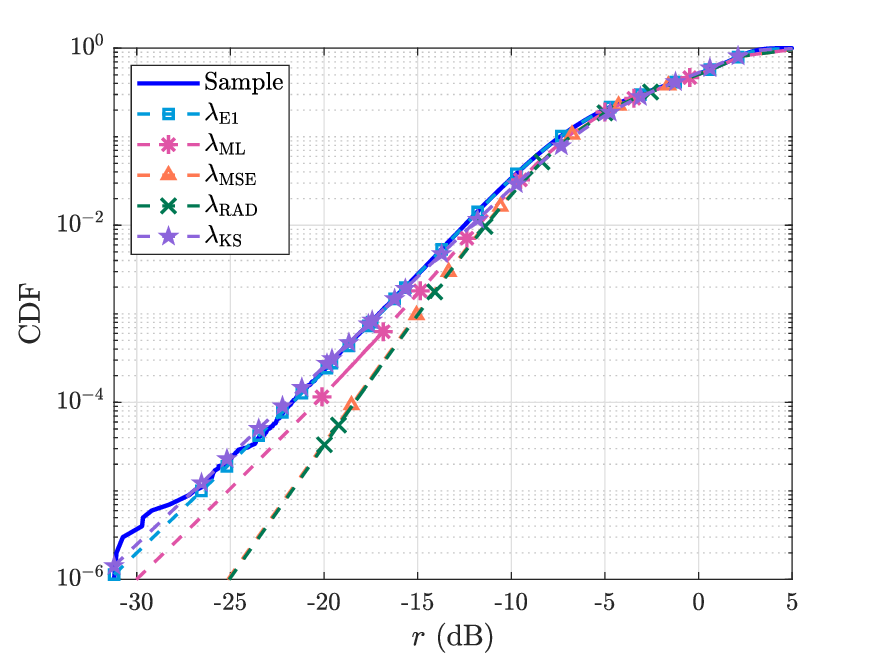}\label{fig:Synth_CDF}}
    \subfigure[Benchmark performance metrics vs $\overline{\gamma}$ (dB). 
    ]
    {\includegraphics[width=0.3\textwidth]{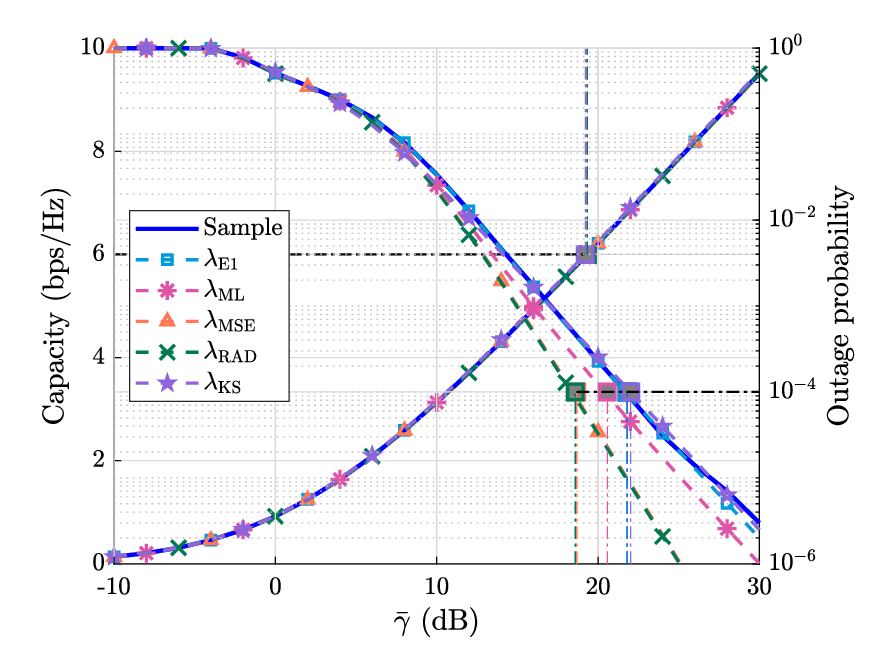}\label{fig:Synth_OP_C}}
    \vspace{-2mm}
    \caption{PDF, CDF and performance evaluation corresponding to E1.}
    \label{fig1Total}
    \vspace{-4mm}
\end{figure*}
\begin{figure*}[ht!]
\vspace{-2mm}
    \centering
    \subfigure[\ac{PDF} vs. $r$.] {\includegraphics[width=0.3\textwidth]{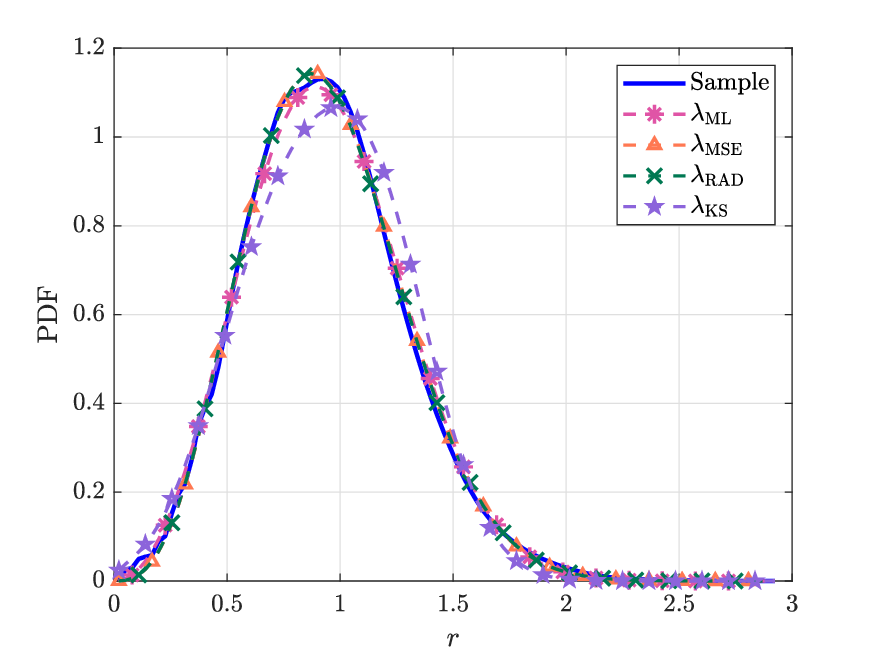} \label{fig:Emp_PDF_Omni}} 
    \subfigure[\ac{CDF} vs. $r$ (dB).]
    {\includegraphics[width=0.3\textwidth]{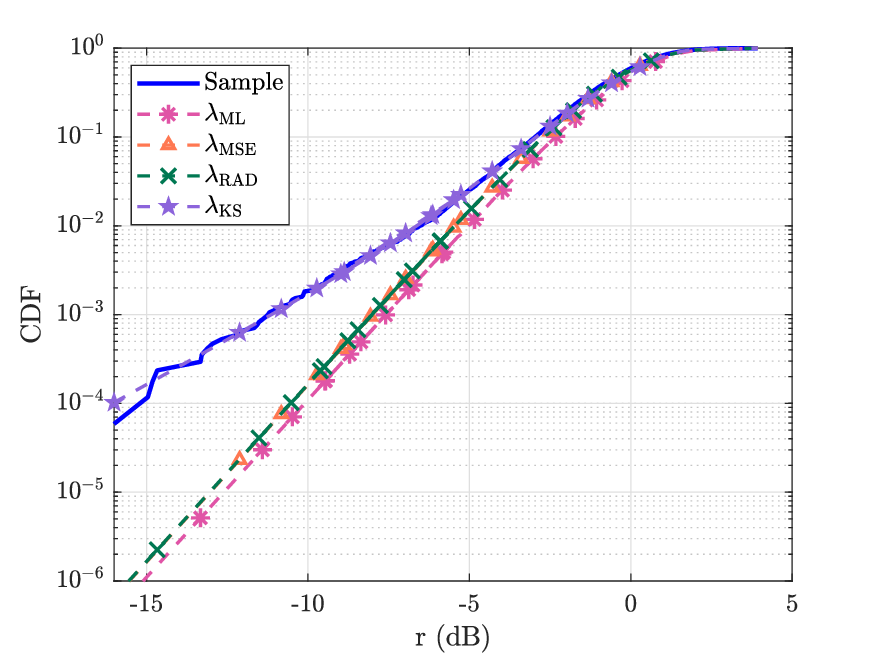}\label{fig:Emp_CDF_Omni}}
    \subfigure[Benchmark performance metrics vs $\overline{\gamma}$ (dB). 
    ] 
    {\includegraphics[width=0.3\textwidth]{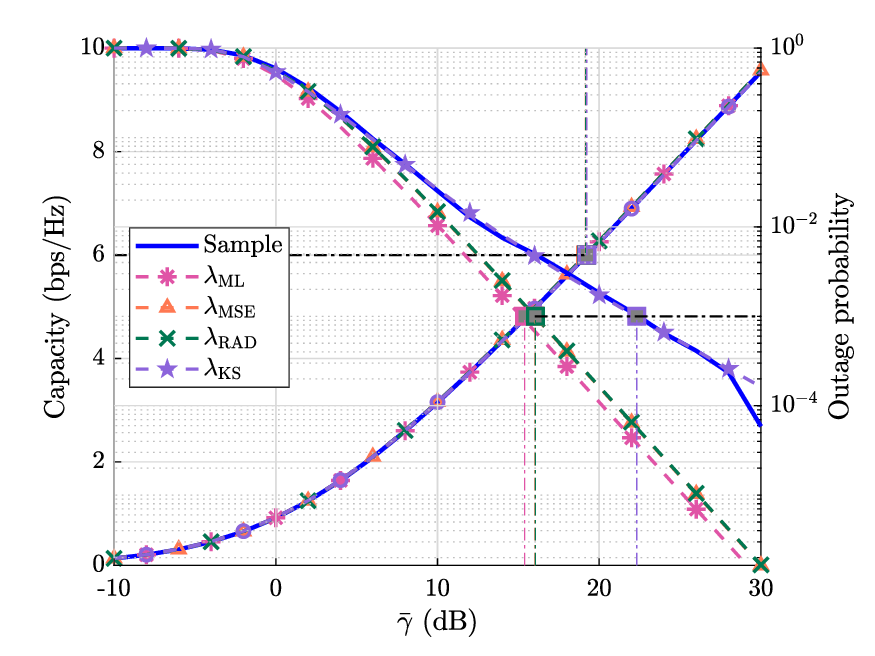}\label{fig:Emp_OP_C_Omni}}
    \vspace{-2mm}
    \caption{PDF, CDF and performance evaluation corresponding to E2. }
    \label{figOmniTotal}
    \vspace{-4mm}
\end{figure*}

\begin{figure}[ht!]
\centering 
\includegraphics[width=.75\columnwidth]{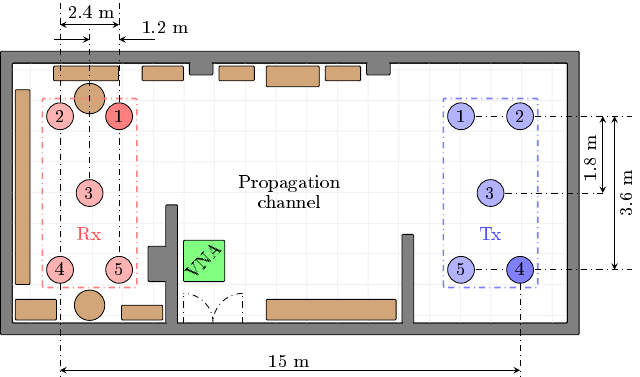}
\caption{\small W-band measurement set-up and environment in E2.}
\label{fig:Setup}
\end{figure}
\vspace{-4mm}
\subsection{Experiment 1}
In Fig. \ref{fig:Synth_PDF}, a set of \ac{PDF}s is represented, including the empirical \ac{PDF} estimated by using \texttt{ksdensity} from sample data $\boldsymbol{\rm r}_{\rm E1}$ of size $n=10^6$, and the \ac{PDF}s obtained by evaluating the set of parameters $\boldsymbol{\rm \lambda}_j$ (listed in Table \ref{tab:GoF_Synthetic_Empirical}). The predefined parameters used to generate synthetic data is $\boldsymbol{\rm \lambda}_{\rm E1}=\left[K_{\rm E1};\Delta_{\rm E1};\mu_{\rm E1}\right]=\left[15, 0.9, 2\right]$, which have been normalized so that $\mathbb{E}\left\{|r|^2\right\}=1$, and the theoretical \ac{PDF} corresponding to $\boldsymbol{\rm \lambda}_{\rm E1}$ is also included for reference purposes. The solution space for the parameter estimation is limited to {$K\in[0.1,45],\Delta\in[0,1],\mu\in[0.1,6]$.}

Visual inspection of Fig. \ref{fig:Synth_PDF} reveals that \ac{MSE}, \textcolor{black}{ML} and \ac{RAD} criteria yield similar performances, being the corresponding \ac{PDF}s virtually indistinguishable from the empirical and theoretical ones. A deeper inspection of parameter values in Table \ref{tab:GoF_Synthetic_Empirical} indicates that parameter estimates minimizing such criteria exhibit some differences compared to $\boldsymbol{\rm \lambda}_{\rm E1}$, i.e. slightly smaller $K$ and larger $\mu$ {\color{black} for the MSE and RAD criteria, and slightly larger $K$ and smaller $\mu$ for the ML criterion}. This is due to the fact that the solution space for the parameters is rather smooth, i.e., similar shapes of the \ac{PDF} are obtained for similar parameter values. Conversely, the shape of the \ac{PDF} corresponding to the set of parameters $\boldsymbol{\rm \lambda}_{\rm KS}$ exhibits a certain mismatch with the sample \ac{PDF}, which is translated into a larger error metric according to the \ac{MSE} and \ac{RAD} criteria, \textcolor{black}{and a lower} $\mathcal{\textcolor{black}{L}}$. 

In Fig. \ref{fig:Synth_CDF}, we now represent the log-\ac{CDF}s corresponding to the same set of parameters as in Fig. \ref{fig:Synth_PDF}. Since the KS criterion pursues that the maximum difference between the empirical and target log-CDFs is minimized, the evaluation of the \ac{CDF} with the set of parameters $\boldsymbol{\rm \lambda}_{\rm KS}$ tightly matches the empirical one (estimated with \texttt{ecdf}) across the entire range of amplitude values -- even better than the \ac{CDF} evaluation with $\boldsymbol{\rm \lambda}_{\rm E1}$. This is due to the fact that the effect of the finite sample is noticed in the left tail of the log-\ac{CDF}, and the fitting procedure is designed to optimize the fit to the sample \ac{CDF}. Interestingly, the CDFs obtained for the sets of parameters $\boldsymbol{\rm \lambda}_{\rm MSE}$ and $\boldsymbol{\rm \lambda}_{\rm RAD}$ largely deviate from the empirical \ac{CDF} in the lower range of amplitude values, precisely the operational range that determines error (and hence outage) events. {\color{black}For the case of $\boldsymbol{\rm \lambda}_{\rm \textcolor{black}{ML}}$, the log-CDF behaves reasonably close to the empirical one, but also some deviation is observed.}

In Fig. \ref{fig:Synth_OP_C}, we analyze the performance of wireless communication systems operating over the sample channel data, and determine whether the different parameter estimation mechanisms turn out being useful to evaluate the benchmark performance metrics defined in Section \ref{Sec:PM}. With regard to the evaluation of the \ac{EC}, we see that all 
\textcolor{black}{fitting} criteria provide excellent results. However, we see that the accuracy in \ac{OP} estimation is only accomplished when using the \ac{KS} criterion. While achieving a remarkably good fit to the empirical \ac{PDF}, this is not translated into a satisfactory tail behavior -- which is essential for \ac{OP} metrics. 
For instance, while a target \ac{OP} value of $10^{-4}$ and $R_{\rm th}=1$ bps/Hz requires an operational \ac{SNR} of roughly $21.8$ dB for the empirical \ac{OP}, the use of \ac{MSE}  and \ac{RAD} criteria ends up estimating an operational \ac{SNR} $\sim 3$ dB smaller {\color{black} ($\sim 1.2$ dB smaller for the case of \textcolor{black}{ML}}; instead, the \ac{KS} criterion has a minor impact ($\sim0.2$ dB loss). This suggests that the specific features of the modified \ac{KS} criterion in \eqref{eq4}, i.e. supremum-based and tail awareness, provide additional robustness for estimating outage-based metrics, even if yielding a larger fitting error \textcolor{black}{according to other criteria}.
\vspace{-3mm}
\subsection{Experiment 2}
This experimental sample corresponds to an indoor measurement campaign in the W-band ($75$-$110$ GHz) as described in Fig. \ref{fig:Setup}. 
For each pair of Tx/Rx positions, a sample of $n=17001$ complex measurements is obtained, corresponding to $S_{21}$ parameter. Tx/Rx are equipped with Eravant omnidirectional antennas \cite{Omni_Antena_Datasheet}, and measurements are performed using the ZVA24 VNA analyzer \cite{VNA_DataSheet} and the ZVA-Z110E \cite{ZVA_DataSheet} converters. 
For exemplary purposes, \ac{GoF} metrics corresponding to the Tx-Rx configuration $4$-$1$ highlighted in Fig. \ref{fig:Setup} are included in Table \ref{tab:GoF_Synthetic_Empirical}, and the fitting results are discussed in the next set of figures for a normalized amplitude envelope. 

For the \ac{PDF} evaluations in Fig. \ref{fig:Emp_PDF_Omni}, we observe an akin behavior as with E1: average-error and \textcolor{black}{likelihood-based } metrics behave similarly and provide similar \ac{PDF} shapes; conversely, the \ac{PDF} obtained from the parameter set minimizing $\epsilon_{\rm KS}$ seems to deviate from the sample, at least in this scale. When inspecting the log-\ac{CDF} evaluations in \ref{fig:Emp_CDF_Omni}, we observe that this latter case provides an excellent fit to the sample \ac{CDF}, whereas those based in \textcolor{black}{likelihood-based} and average-error metrics exhibit a noticeable gap. 

In Fig. \ref{fig:Emp_OP_C_Omni}, we evaluate whether these aspects impact the benchmark performance metrics. Once again, we see that the evaluation of the \ac{EC} metric for each of the parameter sets yield excellent estimates of the \ac{EC} obtained from the empirical dataset. However, for a target \ac{OP} value of $10^{-3}$ and $R_{\rm th} = 1$ bps/Hz, there is only a $\sim 0.2$ dB difference in operational \ac{SNR} between the experimental \ac{OP} and the \ac{KS} approach, while the difference between the experimental \ac{OP} and the average-error \textcolor{black}{and likelihood-based} approaches \textcolor{black}{is larger than} $6.5$ dB.
\vspace{-3mm}
\section{Conclusions}
\label{Sec:Conclusions}
This paper provided a novel look at the interplay between \ac{GoF} metrics for channel fitting and performance analysis measures. Experiments show that many of the widely used \ac{GoF} metrics based on \textcolor{black}{likelihood} or average-error formulations often fail to predict outage events, even under low estimation errors. Instead, the use of \ac{GoF} metrics that (\textit{i}) minimize the maximum instantaneous error; and (\textit{ii}) natively include tail awareness, seem to be more adequate to estimate performance metrics in the ergodic and outage regimes.

\bibliographystyle{ieeetr}
\bibliography{Referencias_ChannelModeling}

\end{document}